%% file: Uravnotezenje_cse.tex
\documentclass[lettersize,journal]{IEEEtran}
\usepackage{amsmath,amsfonts,amsfonts}
\usepackage{algorithmic}
\usepackage{array}
\usepackage[caption=false,font=normalsize,labelfont=sf,textfont=sf]{subfig}
\usepackage{textcomp}
\usepackage{stfloats}
\usepackage{url}
\usepackage{verbatim}
\usepackage{graphicx}
\usepackage{nomencl}
\usepackage{multirow}
\usepackage{array, makecell} 
\usepackage{textcomp}

\usepackage{tikz}
\usepackage{tikzscale}
\usepackage{enumitem}
\usetikzlibrary{calc}
\usetikzlibrary{arrows}
\usetikzlibrary{shapes.geometric}
\usetikzlibrary{arrows.meta,arrows}
\usetikzlibrary{shapes,arrows,circuits,calc,babel}
\UseRawInputEncoding

\def\BibTeX{{\rm B\kern-.05em{\sc i\kern-.025em b}\kern-.08em
    T\kern-.1667em\lower.7ex\hbox{E}\kern-.125emX}}
\usepackage{balance}

\makenomenclature
\usepackage{etoolbox}
\renewcommand\nomgroup[1]{%
  \item[\bfseries
  \ifstrequal{#1}{A}{Sets and Indices}{%
  \ifstrequal{#1}{P}{ Parameters}{%
  \ifstrequal{#1}{V}{Variables}{}}}%
]}

\begin{document}
\title{A Hybrid Renewable-Battery-Electrolyzer Facility under the Single Imbalance Pricing Scheme}
\author{Petra Dra{\v{s}}kovi{\'c}, \emph{Student Member, IEEE}, Ivan Pavi{\'c}, \emph{Member, IEEE}, Karlo \v{S}epetanc, \emph{Student Member, IEEE}, and Hrvoje Pand{\v{z}}i{\'c}, \emph{Senior Member, IEEE}
\thanks{The authors are with the University of Zagreb Faculty of Electrical Engineering and Computing, Croatia.}}


\maketitle
\begin{abstract}
European energy markets are decentralized and entail balance responsibility of each market player. This stresses the importance of imbalance management of renewable energy sources (RES), as the imbalance payments can strongly reduce their profitability. According to the EU Electricity Balancing Guideline, each European transmission system operator should use the single imbalance pricing method which treats both deviation directions the same, no matter if a deviation helps the system or pushes it away from the balance. This paper aims to investigate the behavior of a hybrid facility consisting of an uncontrollable RES, a battery and an electrolyzer under such market setting. The formulated mathematical model of the hybrid facility seeks to maximize  profit in the day-ahead energy market, while minimizing the imbalance costs. Uncertainty of the RES output is captured using stochastic scenarios, while the direction of the power system deviation, relevant for the imbalance pricing, is modeled using a newly proposed robust approach.

Results of the case study indicate that the single imbalance pricing scheme might bring flexible assets to temptation of intentional deviations should they anticipate favorable imbalance prices. 
\end{abstract}

\mbox{}

\input{_shorthands}

\nomenclature[A]{\(\Omega^{t}\) }{Set of time periods}
\nomenclature[A]{\(\Omega^{i}\) }{Set of breakpoints in linearized battery charging curve}
\nomenclature[A]{\(\Omega^{j}\) }{Auxiliary set for linearization of soe }
\nomenclature[A]{\(\Omega^{s}\) }{Set of RES output scenarios }
\nomenclature[P$\lambda$]{$\DAprice $ }{Electricity price in the DA market (\texteuro/MWh)  }
\nomenclature[P$\lambda$]{$\Wprice $ }{Price of water (\texteuro/m$^{\mathrm{3}}$)  }
\nomenclature[P$\lambda$]{$\Hprice $ }{Price of hydrogen (\texteuro/kg)  }
\nomenclature[P$r$]{$\windplanned $ }{Planned production of RES (MW)}
\nomenclature[P$r$]{$\windreal  $ }{Realized production of RES (MW)}
\nomenclature[P$\eta$]{$\eff $ }{Battery roundtrip efficiency }
\nomenclature[P$r$]{$ \R $ }{Coefficients for battery charging curve linearization}
\nomenclature[P$f$]{$ \F $ }{Coefficients for battery charging curve linearization}
\nomenclature[P$p$]{$\pmaxpar $ }{Grid connection power limit (MW)}
\nomenclature[P$\delta$]{$\deltat $ }{Time period (h)}
\nomenclature[P$s$]{$\soemaxpar$ }{Maximum battery capacity (MWh)}
\nomenclature[P$p$]{$\pmax $ }{Installed power of the battery inverter (MW)}
\nomenclature[P$p$]{$\elmax $ }{Maximum power of the electrolyzer (MW)}
\nomenclature[P$\eta$]{$\eleff $ }{Coefficient of water-to-hydrogen conversion (m$^{\mathrm{3}}$/kg)}
\nomenclature[P$\vartheta$]{$\elCoef$ }{Power-to-hydrogen conversion coefficient (MW/kg)}  
\nomenclature[P$\alpha$]{$\alfa$ }{Variable electrolyzer inefficiency coefficient}
\nomenclature[P$\beta$]{$\bet$ }{Fixed electrolyzer inefficiency coefficient}
\nomenclature[P$\zeta$]{$\elmin$ }{Minimum stable operation of the electrolyzer (\%)}
\nomenclature[P$\kappa$]{$\coef$ }{Coefficient for calculating the imbalance price}
\nomenclature[P$\pi$]{$ \pi_{s} $ }{Probability of RES scenario realization}
\nomenclature[P$\Gamma$]{$\Gamma $ }{Uncertainty budget}
\nomenclature[V$x$]{$\binH$ }{Binary variable, 1 if electrolyzer is on, 0 otherwise}
\nomenclature[V$x$]{$\binB$ }{Binary variable, 1 if the battery is participating in balancing of the facility, 0 otherwise}
\nomenclature[V$x$]{$\binDA$ }{Binary variable, 1 if battery is charging, 0 otherwise}
\nomenclature[V$b$]{$\binDual$ }{Indicates the direction of deviation, 1 if the entire system deviates in the unfavourable direction, 0 if the entire system deviates in the favourable direction}
\nomenclature[V$e$]{$\elDA $ }{Electrolyzer power consumption at the DA market (MW)}
\nomenclature[V$\chi$]{$\hyd $ }{Produced hydrogen amount (kg)}
\nomenclature[V$e$]{$\phcrtano$ }{Power consumption of the electrolyzer (MW)}
\nomenclature[V$e$]{$\ph$ }{Actual power required to produce a kilogram of hydrogen (MW)}
\nomenclature[V$e$]{$\elEUplus $ }{Balancing power provided by electrolyzer; positive direction (MW)}
\nomenclature[V$e$]{$\elEUminus $ }{Balancing power provided by electrolyzer; negative direction (MW)}
\nomenclature[V$d$]{$\disEUplus $ }{Balancing power provided by battery discharging; positive direction  (MW)}
\nomenclature[V$d$]{$\disEUminus $ }{Balancing power provided by battery discharging; negative direction  (MW)}
\nomenclature[V$c$]{$\chEUplus $ }{Balancing power provided by battery charging; positive direction (MW)}
\nomenclature[V$c$]{$\chEUminus $ }{Balancing power provided by battery charging; negative direction (MW)}
\nomenclature[V$r$]{$\realv $ }{Part of RES production injected into the grid (MW)}
\nomenclature[V$d$]{$\dev $ }{Deviation of the facility (MWh)}
\nomenclature[V$c$]{$\chDA $ }{Battery charging power at the DA market (MW)}
\nomenclature[V$d$]{$\disDA $ }{Battery discharging power at the DA market (MW)}
\nomenclature[V$r$]{$\real $ }{Realized production of the facility (MWh)}
\nomenclature[V$m$]{$\markpos $ }{Market position of the facility (MWh)}
\nomenclature[V$s$]{$\soes $ }{Battery state-of-energy (MWh)}
\nomenclature[V$s$]{$\soepom $ }{Battery state-of-energy at the segment $j$ (MWh)}
\vspace{-3mm}

\printnomenclature

\vspace{-2mm}

\section{Introduction}
\vspace{-2mm}
\subsection{Motivation}
All major interconnected power systems  are undergoing a transition toward low-carbon electricity generation, spurred by a range of acts such as Fit-for-55 in Europe \cite{fitfor55}, Inflation Reduction Act and Bipartisan Infrastructure Law in the US \cite{amerika1}, \cite{amerika2}, and Climate Solutions Package
 in Australia \cite{Australia}.
The main characteristic of all these acts is a rapid deployment of renewable energy sources (RES), which, besides their obvious advantages, also pose certain challenges. Technical challenges, primarily the intermittency and reduced predictability, manifest as a variety of market challenges. This paper focuses on the European market design, which relies on the principle of balance responsibility, i.e. every market participant is responsible for their imbalances. As a Balance Responsible Party (BRP), each market participant balances its market position or delegates this responsibility to the leader of its balancing group \cite{entsoe}. During the day-ahead (DA) planning or intraday re-planning stage, every BRP has to submit their schedules to the Transmission System Operator (TSO). In real-time, the TSO balances the system by activating reserve resources, and each BRP is charged for own imbalance ex post. There are multiple methods for calculating the imbalance price, but according to Article 55 of the Electricity Balancing Guideline \cite{european2017commission}, each TSO should use the single imbalance pricing method \cite{entsoe2}, which we pursue in this paper. Single imbalance pricing means that both the positive and negative imbalances are treated the same way, regardless if they help or harm the system. 

Coupling RES with batteries and/or electrolyzers provides an opportunity to reduce the imbalance costs as the output of these flexible devices can be controlled. 
In RES-electrolyzer-battery hybrid systems, energy from RES can be used for 1) producing hydrogen via the electrolyzer, 2) storing energy in the battery, 3) injecting it into the power grid. The battery can be used to 1) balance the RES generation, 2) perform arbitrage using either the energy from the RES or from the grid, 3) better distribute the electrolyzer operation. Finally, the electrolyzer operation can be used, besides producing hydrogen, to balance the RES production. This hybrid configuration enables the facility to improve its efficiency by running the devices at higher efficiency rates, by increasing the value of the local renewable energy, and by reducing the imbalance costs.  

\vspace{-2mm}
\subsection{Literature Review}

To attract investors, the initial RES integration was largely driven by support schemes \cite{melliger2021effects}. The first such scheme was the feed-in tariff, which enabled RES to be paid a fixed tariff for the produced energy and, more importantly, relieved them from the imbalance responsibility \cite{banja2017renewables}, \cite{fit}. As RES operators were the only market players not responsible for their imbalances, they were not motivated to plan their output. 
Balancing the RES under the feed-in-tariff was assigned to the TSOs or market operators under their RES balancing groups, e.g. \cite{hrote}. 

The RES operators become BRPs upon replacing the feed-in tariffs with the feed-in premiums \cite{fip} or through direct RES trading in the markets without incentivizing tariffs in the background, equalizing them with all other market participants. 
As the main RES output characteristics are reduced predictability and intermittency, balance responsibility became a major challenge to RES operators.

The literature review section covers three distinctive topics, first the technologies used for imbalance management, second the pricing mechanisms, and third the uncertainty management.

\subsubsection{Technologies Review}

A straightforward solution for covering the production shortfalls is coupling of RES with a controllable generator \cite{4112105}. 
Coupling an RES with energy storage may cover deviations in both directions, but for a limited amount of time. Even though many types of storage can be used to complement an RES plant, e.g. pumped-storage \cite{garcia2008stochastic}, modularity and versatility of battery storage makes it the most common option for daily storage. Fast response time, high energy density, and high efficiency makes batteries suitable for covering short-term fluctuations \cite{haruni2012novel}, while high investment costs, high self-discharge rate, and low specific energy and volume render them inappropriate for long-term storage \cite{raventos2021modeling}. Another solution for RES variability and uncertainty, to which the EU is increasingly inclined, is the diversification of energy vectors through the production of green hydrogen \cite{EU_hydrogen}. This increasingly popular energy vector offers many benefits, such as the provision of fast ancillary services due to quick electrolyzer's ramp-ups \cite{Advisory2020105}, \cite{baumhof2023optimization} and low losses in terms of storage \cite{pavic2022pv}. Paper \cite{pavic2022pv} investigates hybrid facilities taking part in the DA and balancing markets, however, it observes both the RES generation and the imbalance price as deterministic parameters. The presented model does not consider the uncertainty of the power system imbalance direction, which is crucial for determining the imbalance revenue/cost of the hybrid facility. Paper \cite{luth2022electrolysis} assesses the potential of flexibility contribution of electrolyzers on energy islands and accordingly analyzes different market integration strategies. In paper \cite{saretta2023electrolyzer} the authors proposed a mixed-integer linear model that aims to derive an optimal scheduling strategy for an electrolyzer providing frequency containment reserve services in the Nordic synchronous region. On the other hand, the focus of \cite{li2023approach} is on sizing a PV-battery-electrolyzer-fuel cell energy system that utilizes hydrogen as a long-term storage medium and battery as a short-term storage medium to meet the load demand of the buildings at a field lab.

Flexible facilities, consisting of an RES and a flexible technology, besides neutralizing the RES production deviations, are able to increase profit by performing market arbitrage. The simplest example of market arbitrage includes a single, usually the day-ahead, market. An RES can be used to charge a battery during the low-price periods, which can be discharged during the high-price periods \cite{song2016purchase}. More complex examples include two or more electricity markets. A flexible facility can perform arbitrage by trading in consecutive markets, for example, the DA and the intraday (ID) market \cite{vcovic2020optimal} or the DA and the reserve market \cite{soares2016optimal}. 
Besides trading in only electricity markets, flexible facilities that include other energy vectors, in our case hydrogen, can participate in multiple energy/emission markets, for example in the energy, the emissions, and the gas markets \cite{li2013multimarket}, or the energy and the hydrogen markets \cite{miljan2022optimal}, \cite{8086200}. 
Being able to convert between different energy vectors, e.g. electricity-to-hydrogen, enables further reduction of the electricity deviations due to errors in RES output forecasts.
The potential for diversification of energy vectors is still quite unexplored, especially in the context of imbalance management. Most of the existing papers approach the imbalance management of RES facilities by using only one energy vector, i.e. RES and batteries. The potential of conversion between different energy vectors to reduce a hybrid facility's imbalance with respect to the imbalance direction of the power system has not yet been explored, presumably due to low technical efficiency of the electrolyzer as compared to the battery. However, if the hydrogen price is sufficiently high, it can offset high technical losses. Moreover, electrolyzers are highly flexible and can be used to minimize RES uncertainty. This increases the value of the electrolyzer as now it conducts two services: RES balancing and hydrogen sales.

Our work does not consider using electrolyser as energy storage (which should be coupled with a fuel cell and a hydrogen tank in that case), but we use electrolyser as an alternative way to extract renewable energy through hydrogen as an energy vector. The main reason is that the hydrogen storage (transformation of electricity into hydrogen via an electrolyzer and then back to electricity via a fuel cell) suffers from low efficiencies. This is closely aligned with the EU plans for future electrolyser roll out. European Union set goals to install at least 6 GW by 2024, and at least 40 GW by 2030, and to establish transport and develop the backbone of the hydrogen network in Europe \cite{entsog}.

\subsubsection{Imbalance Pricing Mechanisms Review}
In decentralized markets, such as the European market, two distinct electricity imbalance pricing mechanisms can be enforced. Treating both the positive and the negative directions the same way, i.e. the price for negative imbalance and the price for positive imbalance are equal in sign and size, is referred to as the single imbalance pricing. On the other hand, in dual imbalance pricing the price of a negative imbalance is not equal to the price of a positive imbalance in sign and/or size \cite{clo2019effect}.
 
Although according to Article 55 of the Electricity Balancing Guideline \cite{european2017commission}, each TSO in the EU is obliged to use the single imbalance pricing method \cite{entsoe2}, the previous practice in most European energy markets was to use the dual imbalance pricing mechanism. Therefore, in paper \cite{pandvzic2013offering} the authors propose a two-stage stochastic mixed-integer linear program (MILP) that maximizes a virtual power plant's expected profit considering dual imbalance pricing. The virtual power plant participates in both the DA and the balancing markets. The dual imbalance pricing is imposed in a way that the energy can be purchased only at a price higher than in the DA market (up-regulation) and sold at a price lower than in the DA market (down-regulation). In \cite{du2016managing}, the authors also consider dual imbalance pricing for a wind power plant participating in both the DA and the reserve markets. The imbalance is settled at the DA spot price for deviations opposite in sign from the system imbalance, while the deviations of the same sign are settled at the clearing price of the real-time balancing market. Paper \cite{matsumoto2021mitigation} develops a predictive approach for the imbalance volumes and price densities using two-step quantile regressions and derives a new trading optimization for a virtual trader arbitrage position considering dual imbalance pricing. The authors first provide a brief overview of the considered balancing mechanisms since the beginning of the market trading in the Japan Electric Power Exchange in 2005 and conclude that the market has been using the dual imbalance pricing since 2019 to incentivize individual operators to maintain their supply and demand balance instead of taking speculative or gaming positions against the direction of the system deviation.

The authors of \cite{shinde2020optimal} compare three different mechanisms used in practice -- the dual imbalance pricing, the single imbalance pricing, and the single imbalance pricing with spot reversion in the context of an optimal dispatch by a system operator in the balancing markets. The focus is on an analysis of how these three mechanisms help the system operator optimize its balancing market actions considering the RES output. The single imbalance pricing is found to be the most economically efficient. The authors in \cite{marneris2022optimal} 
propose a model that derives the optimal offering strategy of an RES aggregator in the DA energy market and the ancillary services market. The paper also compares single and dual pricing methods, and concludes that the single pricing leaves a headroom for strategic bidding to the aggregator in the form of passive balancing, which translates into higher expected profit. In \cite{bottieau2019very}, the authors examine short-term probabilistic forecasting for risk-aware participation in the single-price imbalance settlement. The developed bi-level model in the upper level maximizes the profit of a market player by identifying its optimal strategic imbalance position, while the lower level simulates the single-price quarter-hourly settlement of energy imbalances. In this paper the imbalance price depends on the price of the reserve  \cite{bottieau2019very}, which is the approach pursued by the European markets.

In addition to the single- and dual-pricing methods, some publications suggest new ones. Papers \cite{schneider2017energy} and \cite{haring2015incentive}, motivated by the negative sides of the currently used mechanisms, propose new imbalance schemes. In \cite{schneider2017energy}, the authors propose an abstract model for market-based imbalance settlement. They claim that such mechanism achieves better trade-offs between the resource efficiency and the bidding accuracy as compared to the alternative penalty price mechanisms. On the other hand, paper \cite{haring2015incentive} proposes an imbalance settlement procedure with an incentive-compatible cost allocation scheme for reserve capacity and deployed energy. The authors claim that their approach guarantees revenue sufficiency for the system operator and provides financial incentives for BRPs to keep their imbalances close to zero. On the other hand, in \cite{morales2012pricing} and \cite{morales2014electricity}, the imbalance price is calculated as the locational marginal price. In \cite{loukatou2021optimal}, the authors assume that in case of a surplus, each market participant is paid additionally by the system operator, but at a lower price than the corresponding wholesale price, while in the case of an under-delivery the participant is penalized. The system deviation direction is not considered. 

Paper \cite{ruben} proposes a novel way to obtain strategic out-of-balance positions for battery storage in a bilevel setup where the lower level entails the balancing energy clearing, while the upper level maximizes the battery storage profit. This approach is highly suitable for market-based imbalance price setting which is taking predominance in Europe, as it proposes a way how to perform arbitrage taking advantage of the imbalance settlement mechanisms. However, this approach is complex and data demanding as the market participants must forecast/estimate both the demand and supply side of the balancing energy market, i.e. both imbalance volumes and balancing energy bids. In our paper, we propose simplified approach where only a ratio of balancing energy prices and day-ahead energy prices is needed. Our approach is also directly applicable in many current imbalance settlement mechanisms in Europe which are still not market-based but regulated.

\textcolor{black}{Since single-imbalance-pricing is a new mechanism in the European markets, thus only a few papers address it. Additionally, papers dealing with single-imbalance-pricing mechanism apply different methods for calculating the imbalance price and do not take into account the direction of deviation of the entire system, which is crucial for determining the price. 
In Croatia, the imbalance price depends on three parameters -- the DA electricity price, the deviation of the entire power system and the coefficient $\coef$ \cite{HOPS_uravnotezenje}. In other words, the imbalance price at a given hour is calculated based on the DA electricity price at the given hour and the deviation direction of the entire power system as follows: 
\begin{itemize}
    \item If the system deviates in the positive direction (i.e. there is a surplus of energy): $price_{imbalance} = (1-\coef) \cdot price_{DA} $
    \item If the system deviates in the negative direction (i.e. there is a lack of energy): $price_{imbalance} = (1+\coef) \cdot price_{DA} $
\end{itemize}
In reality, the imbalance prices are calculated ex post, after the moment of delivery, when the direction of deviation of the entire system is known. Since the deviation direction can not be predicted or known in advance, it is necessary to model the direction of deviation of the entire system. As the imbalance direction of the system is binary, i.e. it can be either positive or negative, we employ a tailor-made robust optimization approach that assumes that the deviation direction of the system is either the same or opposite of the deviation direction of the hybrid facility.}

\textcolor{black}{\subsubsection{Uncertainty Management Review}
Operation of an RES-based facility taking part in energy markets is subject to uncertainties stemming from an uncertain RES output as well as market interaction (including energy and imbalance prices). For the hybrid facility operator to maximized its profits, it is essential to adequately model these uncertainties. Stochastic programming
provides an appropriate mathematical modeling framework both to characterize
the uncertainty and to derive informed decisions \cite{conejo2010decision}. As it characterizes uncertainty using stochastic scenarios, it is essential to have sufficient and reliable historical data. The downside of this method is generally high computational time that increases with the number of scenarios. Stochastic optimization is commonly employed when modeling the uncertainty of RES output. For example, in \cite{conte2017stochastic} the authors use stochastic optimization to model the PV output uncertainty at the day-ahead planning and the real-time control of a PV-battery facility. A benefit of the stochastic optimization is the suitability to employ certain risk measures. The authors in \cite{pandvzic2021managing} employ conditional-value-at-risk to model strategic participation of a battery energy storage in the day-ahead energy as well as reserve
and balancing markets.}

\textcolor{black}{In case historical data are insufficient or unreliable, stochastic optimization may not perform well. In such cases robust optimization might be more suitable as it only requires the upper and lower bounds on the uncertain parameters, without the information on the characterization of uncertainty in between those bounds. Moreover, as opposed to the stochastic optimization, robust optimization is highly effective in terms of computational tractability \cite{pandvzic2015toward}. The objective function in robust models optimizes for the worst realization of uncertainty, however, the robustness of the solution is adjusted using the uncertainty budget. }

\textcolor{black}{The uncertainty related to RES output is internal uncertainty and modern forecasting tools generally work quite well when having historical data that can be used to generate good-quality scenarios. Also, in case of forecast errors, the facility contains flexible units, i.e. battery and electrolyzer, that can be rescheduled if necessary, which is exactly what we aim at with multiple PV output scenarios, as these scenarios affect the expected value of the objective function. On the other hand, the hybrid facility has poor knowledge on the imbalance direction as it is an external parameter. Thus, it makes sense to secure the facility against the worst realization of the system imbalance direction by using the robust approach. The budget of uncertainty can be used to adjust the level of risk aversion.}
\vspace{-2mm}
\subsection{Contribution and Paper Structure}
With respect to the literature review above, we define contribution of this paper as follows:
\begin{enumerate}
    \item A profit maximization model of a hybrid RES-battery-electrolyzer facility taking part in the electricity DA market under the single-imbalance-pricing scheme. The RES is not coupled only with a battery, as in many existing papers, but also with an electrolyzer to demonstrate the potential of hydrogen conversion for preserving the electricity market position. In contrast to the models commonly used in the literature, which do not take into account the reduced possibility of battery charging at higher states of energy, as in \cite{pandvzic2015energy}, and variable efficiency of the electrolyzer, as in \cite{clegg2015integrated}, in this paper both the battery and the electrolyzer are modeled using high-accuracy models \cite{pavic2022pv}, \cite{pandvzic2018accurate}.
    \item The system imbalance price is modeled as a robust subproblem to capture the uncertainty related to the direction of the systems deviation. 
    We develop an approach that considers two outcomes -- the favorable one, where the system deviation direction is in line with the facility's deviation direction, and the unfavorable one, where the system deviation direction is opposite of the facility's deviation direction. Similar results could be achieved using scenario-based uncertainty modeling, as in \cite{bottieau2019very} and \cite{ruben}, however this approach is more complex, data-demanding and harder to be coupled with the hybrid facility's uncertainty.
    Along the uncertain imbalance price direction, we model the facility's deviation using PV scenarios to build a model that maximizes its average earnings but at the same time is secured against huge losses stemming from the imbalance settlement.

\end{enumerate}

Section \ref{Math_model} presents the mathematical formulation of the model, as well as the conversion to a single-level equivalent. Section \ref{Case_study} presents and analyzes the results of a case study for multiple budgets of uncertainty. Section \ref{Conclusion} concludes the paper and highlights the relevant findings.

\vspace{-3mm}
\section{Mathematical Model} \label{Math_model}

\subsection{Objective Function and Robust Transformation}
The proposed mathematical model maximizes the hybrid facility's profit from participating in the electricity DA market considering the imbalance costs, as well as the profit from selling hydrogen produced by the electrolyzer. The facility is a price taker in both the electricity market and the hydrogen market, assuming that all produced quantities can be sold at the market price without affecting the prices, i.e. the facility is a price taker in both markets.

The term in the first row in eq. \eqref{objective_pocetna} is the first-stage decision, i.e. the DA market position of the facility. 
The DA price $\DAprice$ is the hourly probability-weighed expected price. The second row contains the second-stage terms dependent on the RES output scenarios: the revenue from selling hydrogen at a fixed price, the water consumption costs for producing hydrogen, and the electricity imbalance revenue. The imbalance revenue consists of a modified DA price, depending on the entire system's deviation direction. The DA price is either increased or decreased by coefficient $\coef$, depending whether the system deviates in the negative or the positive direction \cite{HOPS_uravnotezenje}. If the entire power system deviates in the positive direction, due to overproduction, the imbalance price is equal to the DA price decreased by coefficient $\coef$. On the other hand, if the system deviates in the negative direction, due to underproduction, the imbalance price is equal to the DA price increased by coefficient $\coef$. 
\vspace{-3mm}
\begin{multline}
     {\max}\sum_{t\in \Omega^{t}} (\DAprice \cdot \markpos)+\\ 
     \!\!\sum_{s\in \Omega^{s}}\! \!\!  \pi_{s}\! \cdot\!   \! \sum_{t\in \Omega^{t}}\! \Bigl( \, \! \Hprice\! \cdot\! \hyd\! -\! \Wprice\!  \cdot\! \eleff \!\cdot \!\hyd \!+ \!\DAprice \!\cdot\! (1 \!\pm\! \coef)\! \cdot\!  \dev\! \Bigr )\,    \label{objective_pocetna} 
\end{multline}

Since there is no correlation between the deviations of the facility and the entire system, the last term in objective function \eqref{objective_pocetna} can in reality be highly favorable or highly unfavorable to the facility operator. To model the uncertainty of the imbalance price, we develop a robust subproblem that aims to select the worse direction of the system deviation with respect to the given uncertainty budget. The expression $\DAprice \cdot (1\pm\coef)  \cdot  \dev$ from objective function \eqref{objective_pocetna} is first separated into two parts: $\DAprice   \cdot  \dev \pm  \DAprice \cdot \coef  \cdot  \dev$, and then the second term is transformed into a robust subproblem. The final objective function that considers the imbalance price deviation in a robust fashion is given in eq. \eqref{objectiveBal1}, where the last three rows represent the robust subproblem. This subproblem decides on the worse direction of the system deviation for the facility. The robust subproblem maximizes over binary variable $\binDual$, which indicates whether $\coef$ is positive or negative. 
\vspace{-3mm}
\begin{equation}
\begin{split}
    &{\max}  \sum_{t\in \Omega^{t}}(\DAprice \cdot \markpos)+\\ &\sum_{s\in \Omega^{s}}   \pi_{s}\! \cdot \! \Bigl( \,  \sum_{t\in \Omega^{t}} ( \Hprice \!\cdot \!\hyd - \Wprice \! \cdot \!\eleff \!\cdot\! \hyd + \DAprice \! \cdot \! \dev ) - \\
    &\!  \underset{\binDual }{\max}  \Bigl\{ \sum_{t\in \Omega^{t}}\! \DAprice  \!\cdot\!  \coef \!\cdot\! |\dev| \hspace{-2.3pt}\cdot\hspace{-2.3pt} (2\! \cdot \! \binDual -1 ) \\ &\text{s.t.} \;\;\; 0\! \leq \!\binDual \!\leq \!1  , \quad \forall{t,\!s} :\!\dualvardva\\
    &\quad\;\;\;\sum_{t} \binDual \leq \Gamma , \quad  \forall{s}:\dualvar 
     \Bigr\} \Bigr)  
       \label{objectiveBal1}
\end{split}
\end{equation}

\begin{figure} [b!]
\vspace{-4mm}
    \centering
    \resizebox*{\columnwidth}{!}
      {
        \input{diagram.tikz}
      }
      \vspace{-5mm}
    \caption{Schematic representation of the interaction between the optimization problem and the robust subproblem}
    \label{shema}
    \vspace{-2mm}
\end{figure}
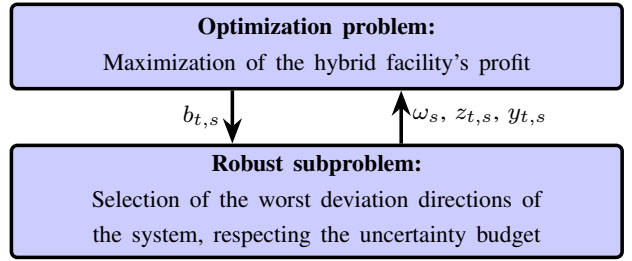

\begin{figure} [b!]
\vspace{-4mm}
    \centering
    \resizebox*{\columnwidth}{!}
      {
        \input{diagram2.tikz}
      }
      \vspace{-8mm}
    \caption{Flowchart of objective function transformations}
    \label{shema2}
\end{figure}
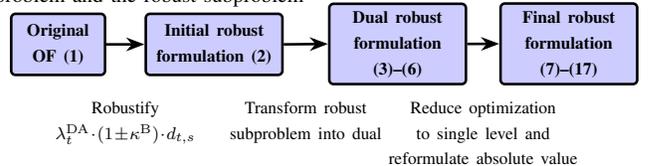

\vspace{-2mm}
Goal of the robust subproblem is to simulate the worst-case realization of uncertainty over the course of the day. However, high level of robustness reduces the expected profit and may cause too conservative bidding decisions. The robustness can be controlled by introducing uncertainty budget $\Gamma$, which defines how many hourly unfavorable deviation directions the model considers. The robust subproblem will select time periods and directions that are least favorable to the facility for $\Gamma$ time periods that have the highest negative consequences to the objective function, while during the remaining time periods the system deviation directions will be considered favorable. This concept is different from the concept of one of the first published papers on robust optimization \cite{bertsimas2004price}, where binary variable $\binDual$ takes value $1$ if an unfavourable deviation occurs, and $0$ if no deviation occurs. However, since power systems practically always deviate in one direction \cite{entsoe_2022}, in our model value $0$ represents a favourable deviation. This is achieved by modifying $\binDual$ from \cite{bertsimas2004price} to $(2\! \cdot \! \binDual \!-\!1)$. The module of the facility's deviation $|\dev|$ from the robust objective function achieves that the binary variable $\binDual$ selects an unfavourable or a favourable system deviation. Also note that binary variable $\binDual$ of the robust subproblem is relaxed to be continuous, as in \cite{bertsimas2004price}, since the relaxation does not change the objective function value, given that the solution of a linear problem can only be found at a vertice of the feasible space. 

A schematic representation of the interaction between the objective function and the robust subproblem is given in Fig. \ref{shema}. Variables $ \omega_{s} $, $ z_{t,s} $, and $ y_{t,s} $ are dual variables of the robust subproblem. The robust subproblem is replaced with its dual so that the objectives of the outer and the inner optimization goals coincide. After replacing the robust subproblem with its dual form, a new objective function \eqref{objectiveDual1} is obtained. Constraints of the robust subproblem are also replaced with their dual constraints \eqref{dual_prvi1}--\eqref{dual_prvi2}.
\vspace{-2mm}
\begin{gather}
    \max\sum_{s\in \Omega^{s}}   \pi_{s} \cdot \Bigl( \, \sum_{t\in \Omega^{t}}(\DAprice  \cdot \dev + \Hprice \cdot \hyd - \Wprice  \cdot \eleff \cdot \hyd  - \nonumber \\
     \dualvardva  +  \DAprice  \cdot \coef \cdot  |\dev|) - \dualvar \cdot \Gamma  \Bigr)
       +  \sum_{t\in \Omega^{t}}(\DAprice \cdot \markpos)   \label{objectiveDual1}\\
    \dualvar + \dualvardva \geq 2 \cdot \DAprice  \cdot \coef \cdot  |\dev|,  \quad \forall  t,s  \label{dual_prvi1}\\
     \dualvar, \dualvardva \geq 0,  \quad \forall  t,s   \label{dual_prvi2} 
\end{gather}

\vspace{-2mm}
The final step is to reformulate the nonlinear absolute value $|\dev|$ using an equivalent MILP formulation. First, the absolute value is replaced by the objective function value of the optimization problem $\min \; \dualvartri \;\; \text{subject to}\; \; \dualvartri \ge \dev \, :\dualfeas \,; \;\; \dualvartri \ge -\dev\,:\dualfeasdva$, which is transformed into a system of equations using the KKT conditions, resulting in equations \eqref{dual_drugi11}--\eqref{dual_drugi2}. The final MILP equivalent of objective function \eqref{objectiveBal1} is given by objective function \eqref{objectiveDual2} and constraints \eqref{dual_drugi1}-\eqref{dual_drugi2}. The flowchart of transformations from the original objective function to the final robust formulation is depicted in Fig. \ref{shema2}.
\vspace{-3mm}
\begin{multline}
    \max\sum_{s\in \Omega^{s}} \!  \pi_{s}\! \cdot\! \Bigl(  \sum_{t\in \Omega^{t}}(\DAprice \! \cdot\! \dev + \Hprice\! \cdot\! \hyd - \Wprice \! \cdot\! \eleff\! \cdot\! \hyd  - \\
     \dualvardva  +  \DAprice  \cdot \coef \cdot  \dualvartri) - \dualvar \cdot \Gamma  \Bigl)
       +  \sum_{t\in \Omega^{t}}(\DAprice \cdot \markpos)   \label{objectiveDual2}
\end{multline}
\vspace{-4mm}
subject to:
\begin{gather}
    \dualvar + \dualvardva \geq 2 \cdot \DAprice  \cdot \coef \cdot \dualvartri,  \quad \forall  t,s  \label{dual_drugi1} \\
    \dualvar, \dualvardva \geq 0,  \quad \forall  t,s  \\
     \dualvartri \geq \dev,  \quad \forall  t,s  \label{dual_drugi11}\\
     \dualvartri \geq - \dev,  \quad \forall  t,s  \\
     1 - \dualfeas - \dualfeasdva = 0,  \quad \forall  t \\
     \dev - \dualvartri \geq - (1 - \binDualFeas) \cdot \bigM,  \quad \forall  t,s  \\     
      \dualfeas  \leq \binDualFeas \cdot \bigM,  \quad \forall  t \\
      -\dev - \dualvartri \geq - (1 - \binDualFeasdva) \cdot \bigM,  \quad \forall  t,s  \\     
      \dualfeasdva  \leq \binDualFeasdva \cdot \bigM,  \quad \forall  t \label{dual_drugi2} 
\end{gather}

\vspace{-4mm}
\subsection{Constraints}

Deviation of the hybrid facility, given by eq. \eqref{eq. deviation}, is defined as a difference between the realized output and the market position. The realized output depends on the actual output of the RES, the charging and discharging DA schedule of the battery, the DA scheduled consumption of the electrolyzer, and the balancing power provided by the battery and the electrolyzer, as defined in eq. \eqref{eq. achieved}. Since both the battery and the electrolyzer can balance the facility in both directions, we distinguish variables with plus and minus in the superscript. Positive sign indicates an increase in charging, discharging, or electrolyzer consumption, while negative sign indicates a decrease. The market position in eq. \eqref{eq. marketposition} depends on the planned production of the RES, the battery's charging/discharging schedule in the DA market, and the electrolyzer's consumption in the DA market. Eq. \eqref{eq. vari} indicates that the utilized renewable output has to be lower or equal to the available output, allowing for curtailment.
\vspace{-2mm}
\begin{gather}
    \dev = \real - \markpos, \quad \forall t,s \label{eq. deviation} \\
    \real \! = \! 
    (\realv \! + \!  \disDA \! - \!  \chDA \! - \! \elDA \!  +  \!  \disEUplus \! - \disEUminus - \nonumber \\
    \chEUplus + \chEUminus \!   - \!  \elEUplus \! + \elEUminus ) \! \cdot \! \deltat, \quad \forall t,s \label{eq. achieved} \\
    \markpos = (\windplanned + \disDA - \chDA - \elDA) \cdot \deltat, \quad \forall t \label{eq. marketposition} \\
    \realv \leq \windreal ,  \quad \forall  t,s  \label{eq. vari} 
\end{gather}

\vspace{-1mm}
Constraints \eqref{eq.ex1}--\eqref{eq.ex2} are grid exchange constraints enforced at both the balancing and the DA stages. 
\vspace{-1mm}
\begin{gather}
    - \pmaxpar \leq \real  \leq \! \pmaxpar \! ,\!  \quad \forall t, s \label{eq.ex1} \\
    - \pmaxpar \leq \markpos \leq \pmaxpar, \quad \forall t \label{eq.ex2}
\end{gather}

\vspace{-1mm}
Constraints \eqref{bat1}--\eqref{bat_zadnja} model the battery. An accurate charging model of the battery that considers the reduced charging ability at high state-of-energy using a piecewise linear approximation is adapted from \cite{pandvzic2018accurate}. Battery state-of-energy is tracked in eq. \eqref{bat1} and considers the DA charging/discharging variables, as well as their up and down balancing counterparts per scenario. State-of-energy in each scenario is limited to the battery available net capacity in constraint \eqref{soemax}. Eq. \eqref{bat_pom} calculates the state-of-energy per scenario using piecewise linear parts, which are limited in constraint \eqref{bat_soej}. Constraint \eqref{bat_soepom} is used to relate the battery charging ability to its state-of-energy. More details on this accurate charging model can be found in \cite{pandvzic2018accurate}. Eqs. \eqref{eq.disEU}--\eqref{eq.chDA} prevent simultaneous charging and discharging of the battery at the balancing and the DA stages. Constraints \eqref{bal_bat1} and \eqref{bal_bat3} prevent that the combined DA and positive balancing charging/discharging power exceed the maximum battery power, while constraints \eqref{bal_bat2} and \eqref{bat_zadnja} model the same for the combination of the DA and the negative balancing charging/discharging power. Finally, the nonegativity is imposed in eq. \eqref{nenegbat}.
\vspace{-2mm}
\begin{gather}
     \soes = \soets + \deltat \cdot \eff \cdot (\chDA + \chEUplus - \chEUminus) - \nonumber \\
    \deltat \cdot \frac{1}{\eff} \cdot (\disDA + \disEUplus - \disEUminus) , \quad \forall t,s \label{bat1}  \\
    \soes \leq \soemax , \quad \forall t, s \label{soemax} \\
    \soes = \sum_{j} \soepom  , \quad \forall t, s \label{bat_pom}\\
      \soepom \leq \sum_{i} (\Rplusi - \R) \cdot \soemax  , \quad \forall t,s,j \label{bat_soej}\\
    \deltat \cdot \eff \cdot  (\chDA + \chEUplus - \chEUminus) \leq \Fjedan \cdot \soemax  -  \nonumber \\
    \sum_{i \geq 2} \sum_{j=i-1} \soepomt \cdot \frac{\Fminusi - \F}{\R - \Rminusi}, \quad \forall t>1, s \label{bat_soepom} \\
    \chEUplus + \disEUminus \leq (1- \binB ) \cdot 2 \cdot \pmax, \quad \forall t, s \label{eq.disEU}\\
        \chEUminus + \disEUplus \leq \binB \cdot 2  \cdot \pmax, \quad \forall t, s \label{eq.chEU} \\
            \disDA \leq (1-\binDA ) \cdot \pmax, \quad \forall t \label{eq.disDA} \\
                \chDA \leq \binDA \cdot \pmax, \quad \forall t \label{eq.chDA} \\
                    \chEUplus + \chDA \leq \pmax , \quad \forall t,s \label{bal_bat1} \\
        \chEUminus \leq \chDA , \quad \forall t, s \label{bal_bat2}\\
        \disEUplus + \disDA \leq \pmax , \quad \forall t, s \label{bal_bat3}
\end{gather}
\vspace{-4mm}

\vspace{-3mm}
\begin{gather}
    \disEUminus \leq \disDA  , \quad \forall t, s \label{bat_zadnja}\\
    \soes,\soepom, \chDA,\chEUminus,\chEUplus, \nonumber \\
    \disDA, \disEUminus,\disEUplus \geq 0, \quad \!\!\! \forall  t,s \label{nenegbat}
\end{gather}

The electrolyzer model uses three electrolyzer variables in eq. \eqref{elektrolizator7} in the same way as in the battery model. Constraint \eqref{elektrolizator1} prevents the electrolyzer from operating at a power lower than the technical minimum or higher than the maximum power. The amount of produced hydrogen is determined in eq. \eqref{elektrolizator5} based on the net power used to produce hydrogen. The net power accounts for the electrolyzer efficiency, which reduces as the electrolyzer power increases, As modeled in eq. \eqref{elektrolizator6}. The nonnegativity is imposed in constraint \eqref{nenegel}, while constraint \eqref{nenegel1} prevents the DA and balancing variables from exceeding the maximum power of the electrolyzer.  Constraint \eqref{elektrolizator8} prevents the combined DA and positive balancing consumption to exceed the maximum electrolyzer power, while constraint \eqref{elektrolizator9} prevents the combined DA and negative balancing consumption to fall below zero.

\vspace{-3mm}
\begin{gather}
     \ph =  \elDA + \elEUplus - \elEUminus , \quad \!\!\! \forall  t,s 	 \label{elektrolizator7} 
      \\
     \elmin \cdot \elmax \cdot \binH\leq \ph  \leq  \elmax \cdot \binH  , \quad \!\!\! \forall  t  \label{elektrolizator1}\\
     \phcrtano = \hyd \cdot \elCoef , \quad \!\!\! \forall  t,s  \label{elektrolizator5} \\
     \phcrtano = \alpha \cdot \ph + \beta \cdot \elmax \cdot \binH, \quad \!\!\! \forall  t,s 	 \label{elektrolizator6}\\
      \elDA, \elEUminus, \elEUplus \geq 0, \quad \!\!\! \forall  t,s \label{nenegel} \\
      \elDA, \elEUminus, \elEUplus \leq \elmax, \quad \!\!\! \forall  t,s \label{nenegel1} \\
      \elDA + \elEUplus \leq \elmax , \quad \!\!\! \forall  t,s 	 \label{elektrolizator8} 
      \\
      \elDA - \elEUminus \geq 0 , \quad \!\!\! \forall  t,s 	 \label{elektrolizator9} 
\end{gather}
\vspace{-10mm}

\section{Case Study and Results} \label{Case_study}

\subsection{Input Data}
\begin{figure}[t!]
\centerline{\includegraphics[width= .5\textwidth]{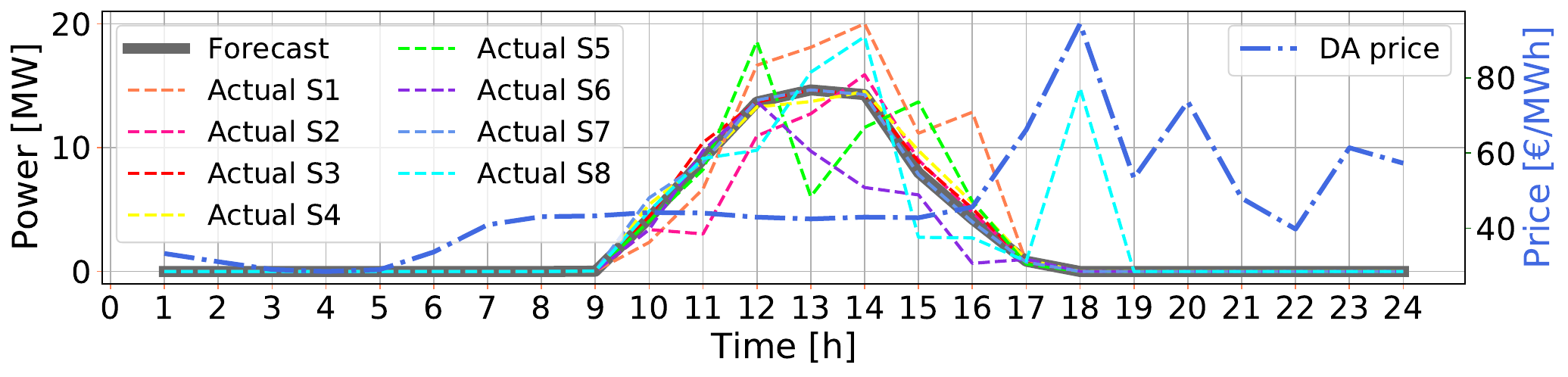}}
\vspace{-4mm}
\caption{Input data -- forecast and actual RES production as well as the DA electricity prices }
\label{fig:input}
\vspace{-4mm}\end{figure}

In the case study we consider a 20 MW PV plant and generate its production curves. The forecast production curve was generated using the PVSol tool \cite{PVSOL}, while eight realization scenarios were generated by the Solcast tool \cite{SOLCast}. 
The forecast curve and the eight realization curves for the day considered in the case study were generated for a location in the southern part of Croatia and are presented in Fig. \ref{fig:input}. The highest positive deviation between the forecast and the actual production is 14.80 MW, while the highest negative deviation is -8.61 MW. Fig. \ref{fig:input} also shows the DA electricity prices used in the case study. The prices are lower in the first part of the day, exhibiting a significant increase in the second part. The lowest price is observed in hour 4 (28.5 \texteuro/MWh), while the highest price of 94.43 \texteuro/MWh is reached in hour 18. 

The battery capacity is 5 MWh, its rated power is 5 MW and roundtrip efficiency is 0.92. State-of-energy at the beginning is set to zero. The electrolyzer maximum power is 5 MW, while its operating minimum is 10\%, i.e. 0.5 MW. Coefficients $\elCoef$, $\alpha$ and $\beta$ are set to 0.0394, 0.689 and 0.011, respectively. As there is no operational hydrogen market, we use a fixed hydrogen price of 2 \texteuro/kg \cite{PWC}, while the cost of water to produce hydrogen is 0.397 \texteuro/m$^{\mathrm{3}}$. 
The grid connection power limit is set to 20 MW. The coefficient for calculating the imbalance price $\coef$ is 0.4 \cite{HOPS_uravnotezenje}.

\begingroup
\setlength{\tabcolsep}{4.0pt} 
\begin{table}[b!]
\vspace{-4mm}
\begin{center}
\caption{Total expected profit (consisting of the electricity DA trading revenue and the expected hydrogen  and imbalance revenues) and real total profit (based on the real imbalance revenue) of the facility for all values of $\Gamma$  }
\label{tab1}
\vspace{-4mm}
\begin{tabular}{| >{\bfseries}c|  >{\bfseries}c  c  c  c |  >{\bfseries}c  c|}
\hline
 & Total  & Electricity   & Expected  & Expected & Real & Real\\

\multirow{2}{*}{$\Gamma$}  & expected  & DA  trad-   &    hydrogen  &  imbalance    &  total & imbalance \\

   &  profit   &   ing  revenue  &    trading rev-   &  revenue  & profit &  revenue\\

   &(\texteuro)    &     (\texteuro)  &     enue (\texteuro) &   (\texteuro) &   (\texteuro) & (\texteuro)\\

\hline
0 & 9010 & -2138 & 1797 & 9350  & 2729 & 3070\\
1 & 8010 & -1546  & 1795 & 7761  & 2804 & 2555\\
2 & 7393 & -1169 & 1720 & 6841  & 2854 & 2303\\
3 & 6841 & -978 & 1677 & 6142  &  2890 & 2191\\
4 & 6330  & -745 & 1623 & 5452  & 2936 & 2057 \\
5 & 5851  & -638 & 1593 & 4897 & 2963 & 2008 \\
6 & 5432 & -1505 & 1099 & 5838  & 3377 & 3784\\
7 & 5047 & -1496 & 1099  & 5444 & 3379 & 3776 \\
8 & 4677  & -2037 & 1006 & 5708 & 3397 & 4428 \\
9 & 4331 & -2484 & 974 & 5841  & 3313 & 4823\\
10 & 4024  & 343 & 1305  & 2376  & 3423 & 1776\\
11 & 3759  & 1008 & 1305 & 1446  & 3490 & 1177\\
12 & 3548  & 2054 & 1205 & 289  & 3138 & -120\\
13 & 3486 & 2532  & 1252  & -299 & 3552 & -232\\
14 & 3469 & 2297 & 1287  & -115 & 3582 & -2 \\
15 & 3467  & 2307 & 1291 & -131  & 3579 & -19\\
16 & 3467 & 2307 & 1291 & -131  & 3579 & -19\\
17 & 3467 & 2307 & 1291 & -131 & 3579 & -19 \\
18 & 3467 & 2307 & 1291 & -131  & 3579 &-19\\
19 & 3467 & 2307 & 1291 & -131  &  3579 &-19\\
20 & 3467  & 2307 & 1291 & -131  & 3579 &-19\\
21 & 3467 & 2307 & 1291 & -131  & 3579 &-19\\
22 & 3467 & 2307 & 1291  & -131  & 3579 & -19\\
23 & 3467  & 2307 & 1291 & -131  & 3579 & -19\\
24 & 3467  & 2307 & 1291 & -131 & 3579 &-19 \\
\hline 
\end{tabular}
\end{center}
\vspace{-3mm}
\end{table}
\endgroup

\vspace{-3mm}
\subsection{Results}
\subsubsection{Economic Analysis}
Table \ref{tab1} shows the expected total profit, i.e. the objective function value, which is made of the electricity DA trading revenue (this is a first-stage variable), the expected hydrogen trading revenue (this is a second-stage variable and is subject to the RES realization), and the expected imbalance revenue for all $\Gamma$ values. The highest total profit is expected for $\Gamma=0$, which assumes favourable system deviations at all hours. Most of the profit comes from the imbalance revenue, while the hydrogen trading revenue is app. five times lower. Depending on the scenario realization, the imbalance stage for $\Gamma=0$ will result in a revenue in between 8308 \texteuro $\,$ and 10320 \texteuro, with an expected value of 9350 \texteuro. Increasing $\Gamma$ to 1 improves the electricity DA revenue and slightly reduces the hydrogen trading revenue. However, it has a strong negative impact on the expected imbalance revenue, whose expectation is now reduced to 7761 \texteuro. This is because the robust subproblem seeks to identify the hour with the highest negative impact on the objective function to change the system deviation direction. As the value of $\Gamma$ further increases, the total expected profit monotonically decreases due to a reduction in the expected imbalance revenue, which is only partially compensated by an increased electricity DA trading revenue and the expected hydrogen trading revenue. For $\Gamma=15$ and above, the expected total profit and facility operation stabilizes at the total expected profit 3467 \texteuro.

As the robust model optimizes the decisions only for the worst-case realization of uncertainty, it is not evident which value of $\Gamma$ would perform best in reality, i.e. in case of materialization of any other but the worst-case uncertainty. Thus, the obtained schedule is tested against thirty scenarios of real systems deviation directions (thirty days within one month taken from \cite{Transparency}). The average values of thirty different real imbalance revenues for each value of $\Gamma$ are listed in the last column of Table \ref{tab1}, which are used to calculate the real total profit in the penultimate column. The highest total profit, considering the real data on the power system deviation directions, is attained for $\Gamma=14$, while the lowest profit would be realized for $\Gamma=0$. Although the imbalance revenue for $\Gamma=14$ is slightly negative, the other two revenues are higher than in other cases with higher imbalance revenues, and therefore the real total profit is the highest. On the other hand, in case $\Gamma=0$ the real imbalance revenue is high, due to a highly aggressive and perilous bidding strategy that aims to take advantage of the imbalance price higher than the DA prices. However, due to real deviation directions, the imbalance revenue is not high enough to compensate for the deficit caused by the negative electricity DA trading revenue. This results in the lowest real total profit, indicating that the overall results are sensitive to the chosen uncertainty budget.

\subsubsection{Impact of Different Technologies}

\begin{figure}[t!]
\centerline{\includegraphics[width= .5\textwidth]{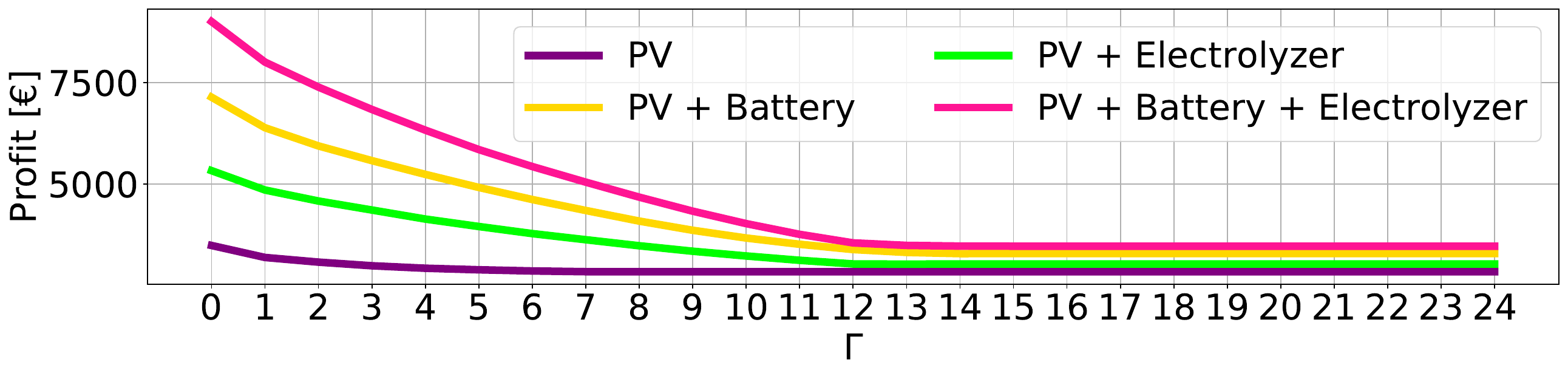}}
\vspace{-4mm}
\caption{The expected profits in case of only PV, PV + electrolyzer, PV + battery, and  PV + electrolyzer + battery for different $\Gamma$ values}
\label{fig:profiti}
\vspace{-4mm}\end{figure}

In this part of the case study we compare the expected profits for different combinations of technologies within the hybrid facility. Fig. \ref{fig:profiti} indicates the expected profits in the case of only the PV, PV+electrolyzer, PV+battery, as well as all three technologies. As expected, the highest expected profit occurs in the case of coupling all three technologies, when the facility has the widest possibility of imbalance management, while the lowest profit occurs in the case of using only the PV. 
Although in all four cases the expected profit decreases as gamma increases, with an increase in the number of coupled technologies, i.e. an increase in the imbalance management capacity, the expected profit curve is characterized by a steeper decline in expected profits. For example, in the case of using only the PV, the expected profit for $\Gamma$=0 is only 22,9\% higher than for $\Gamma$=24, while in the PV+electrolyzer+battery case, the expected profit for $\Gamma$=0 is 159,9\% higher than for $\Gamma$=24.
Even though the nominal capacity of both the battery and the electrolyzer is the same, i.e. 5 MW, the main reason for the expected profit difference lies in the fact that the battery has two times higher capacity in terms of the imbalance management, as it can both charge and discharge at the given power rating, than the electrolyzer of the same nominal capacity. However, the electrolyzer provides a significant increase in the profitability of the hybrid facility, indicating its relevance in the overall income, even for relatively low price of hydrogen considered in this case study of 2 \texteuro/kg.

\subsubsection{Analysis of the First-Stage -- DA Market Positioning}
\begin{figure}[t!]
\centerline{\includegraphics[width= .5\textwidth]{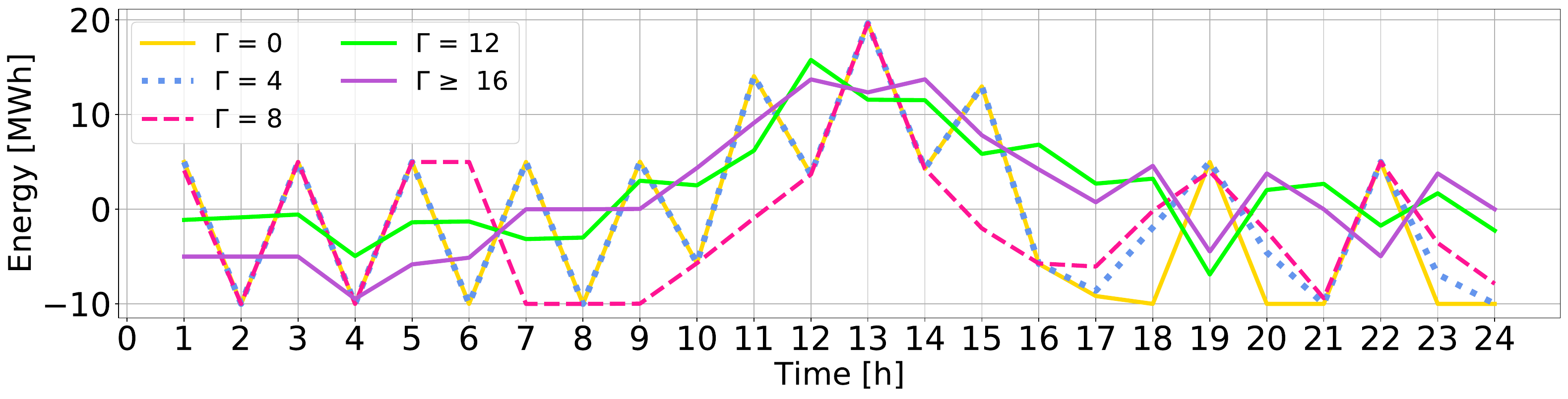}}
\vspace{-4mm}
\caption{DA market position of the facility for different $\Gamma$ values}
\label{fig:marketpos}
\vspace{-4mm}
\end{figure}

To further investigate the results and behavior of the proposed model, Fig. \ref{fig:marketpos} shows the facility's DA market position, i.e. the first-stage decisions, depending on the uncertainty budget. For low values of $\Gamma$ the DA market positioning curve greatly differs from the forecast production curve as the model seeks to perform arbitrage between the DA and the imbalance price. On the other hand, for higher values of $\Gamma$ the DA market positioning curve follows the forecast production curve. This is best seen during hours 7-9, when the RES forecast is zero and the DA prices are not attractive enough for performing temporal arbitrage, so the market position is also zero. For high $\Gamma$ values the model behaves conservatively as it expects to be punished for its deviations in real time. However, for low $\Gamma$ values, the model is confident that the system deviation direction will be favorable to the facility. 
As demonstrated by the results in the previous subsection (see Table I), such aggressive and gambling behavior in reality results in much lower than expected imbalance revenue, resulting in lower overall profit.

\subsubsection{Analysis of the Second-Stage -- Deviations and Scenario Realizations}

\begin{figure*}[t!]
\centerline{\includegraphics[width= \textwidth]{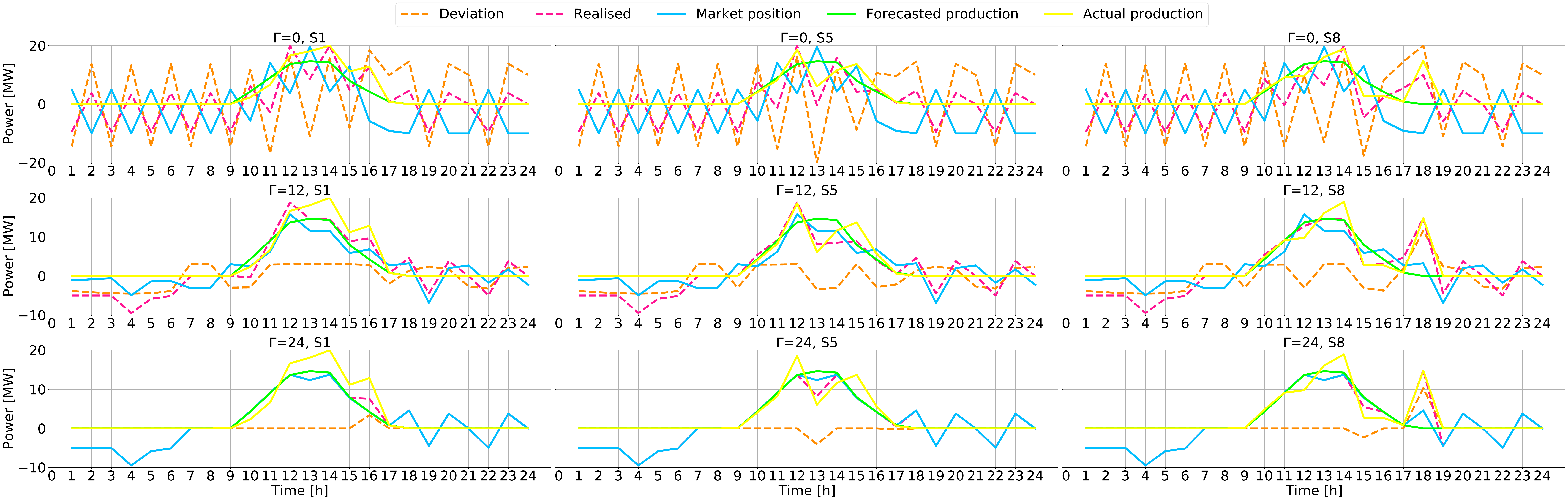}}
\vspace{-2mm}
\caption{Forecast and actual PV production, DA market position, realization and deviation of the facility for $\Gamma=0$, $\Gamma=12$ and $\Gamma=24$ and for scenarios S1, S5, and S8 }
\label{fig:s158}
\vspace{-4mm}
\end{figure*}

Fig. \ref{fig:s158} provides a detailed representation of the forecast and actual PV production, the DA market position, and the deviation for scenarios S1 (actual output generally higher than the forecast), S5 (actual output in some hours higher, in some hours lower than the forecast) and S8 (unforeseen high output in hour 18) for $\Gamma $ values 0, 12 and 24. 
For $\Gamma=0$ the model sets the DA market position in a way to produce the highest possible deviation in real time. Such opposite position enables a 40\% profit at the imbalance settlement as compared to the DA market price (note that $\coef$ in this case study is set to 0.4). In hour 12, when the DA market position is relatively low, the hybrid facility seeks to increase the realization as much as possible. Although in scenarios S1 and S5 the actual production (the yellow curve) is higher than the forecast (the green curve), thanks to the battery, the overall facility realization (the dashed magenta curve) reaches the maximum value (20 MW) to maximize the deviation (the dashed orange curve). On the other hand, in scenario S8, the actual production in hour 12 is lower than the forecast and, despite discharging the battery, the deviation is significantly lower than in scenarios S1 and S5. On the contrary, in hour 13, when the facility is positioned at the maximum value, the realization is minimized to maximize the negative deviation. The maximum deviation is achieved in scenario S5 due to a low actual PV production and high consumption of the battery and the electrolyzer. In scenarios S1 and S8, the actual PV production is higher than the forecast, so the deviations are not as severe as in S5.

With the increase of $\Gamma$, the realization curve is closer to the DA market position curve. In scenario S1, for $\Gamma=12$, it even takes the form of the DA market position curve but is translated upward. In scenarios S5 and S8 the realization curve is more similar to the actual production curve because the capacity of the battery and the electrolyzer are not sufficient to reduce/increase the actual production to the positioning level. In most hours, the facility deviates, but the deviations are much lower than in the previous case.

Finally, for $\Gamma=24$ in most hours the realization curve overlaps with the DA market position curve. Deviations only occur in hours when the battery and the electrolyzer are unable to compensate for the difference between the position and the actual production. In scenario S1, deviations only occur in hours 10 and 16, in scenario S5 in hour 13, and in scenario S8 in hours 15 and 18. 

\subsubsection{Battery Operation Analysis}

\begin{figure}[t!]
\centerline{\includegraphics[width= .5\textwidth]{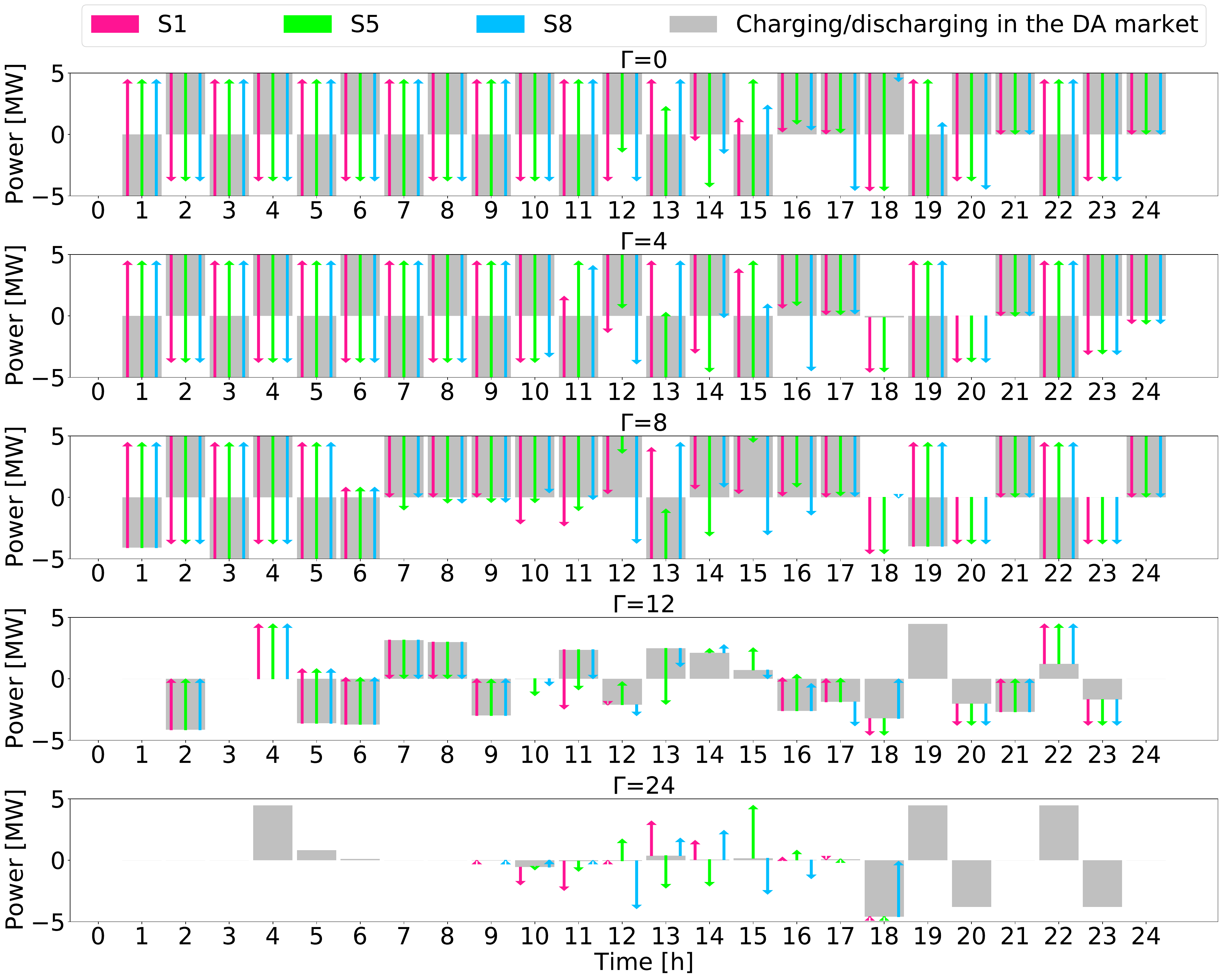}}
\vspace{-3mm}
\caption{Battery DA schedule and balancing actions for different $\Gamma$ values}
\label{fig:battery}
\vspace{-4mm}
\end{figure}

\begin{figure}[t!]
\centerline{\includegraphics[width= .5\textwidth]{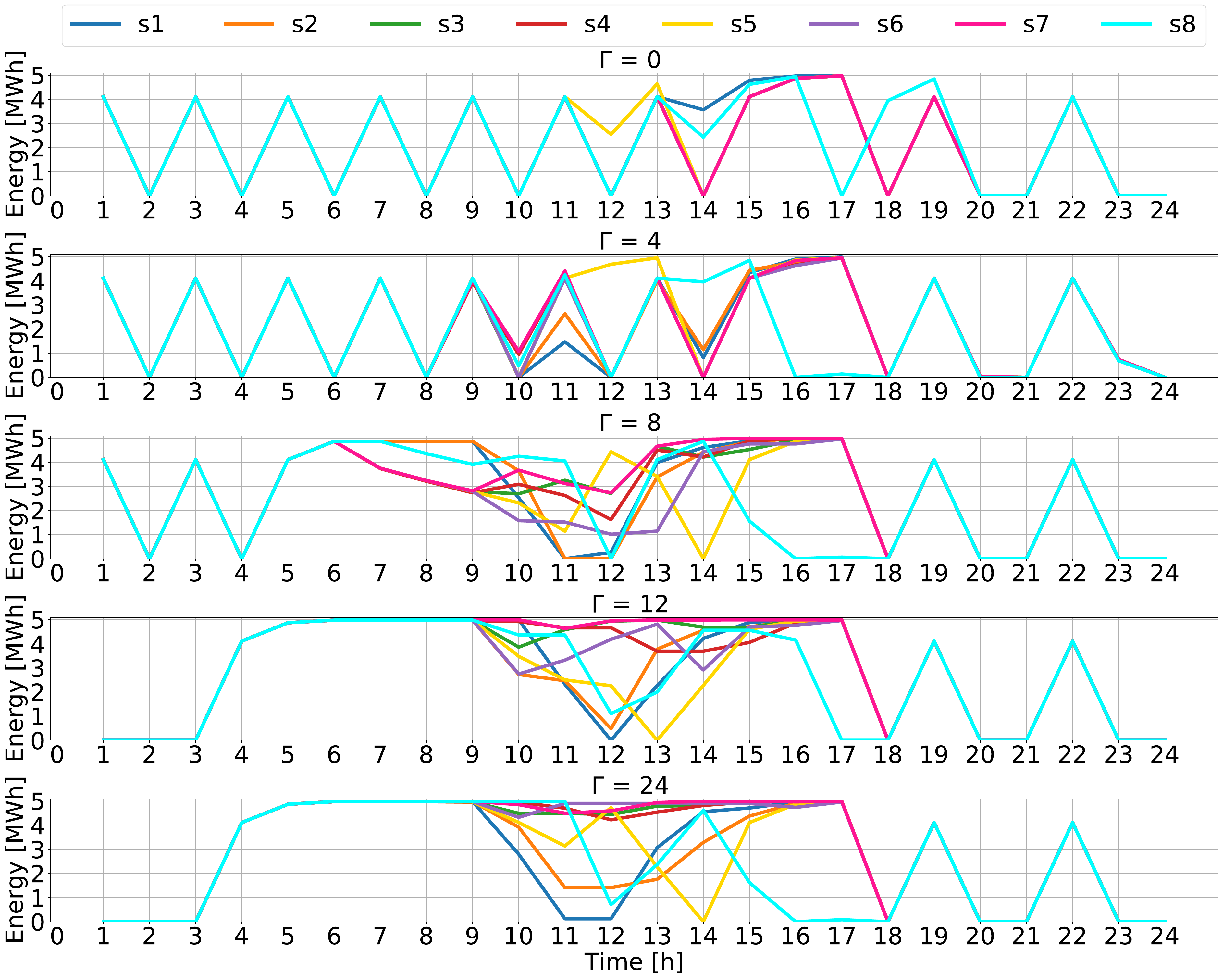}}
\vspace{-3mm}
\caption{Battery state-of-energy for different $\Gamma$ values for each scenario}
\label{fig:soe}
\vspace{-4mm}
\end{figure}

Fig. \ref{fig:battery} presents the battery DA market position (thick grey columns) for different $\Gamma$ values, as well as the balancing power per scenarios S1, S5, and S8 (arrows),

while Fig. \ref{fig:soe} shows the actual battery state-of-energy per scenario (sum of the battery DA position and the balancing power provided by the battery). For low values of $\Gamma$ the battery is dominantly used for performing arbitrage between the DA market and the imbalance settlement, while for high values of $\Gamma$ it is used for intertemporal arbitrage within the DA market. In both cases, it is used to balance the PV output. For $\Gamma=0$ the facility positions to supply 5 MW to the grid in hour 1 (see Fig. \ref{fig:marketpos}). Although there is no planned PV production during this hour and the battery is empty, the facility plans to discharge 5 MW from the battery to the grid at the DA stage. This is because $\Gamma=0$ assumes the system will deviate in a favorable direction, earning the facility $\DAprice \cdot 5$ MW in the DA market, while paying only $0.6 \cdot \DAprice \cdot 5$ MW for the caused imbalance, resulting in $0.4 \cdot \DAprice \cdot 5$ MW profit without actual discharging or delivering energy. Additionally, it is profitable for the facility to actually charge at a 40\% cheaper price than in the DA market. This is confirmed in Fig. \ref{fig:soe}, which shows that even though the battery is positioned to discharge in the DA, in reality it is charged to 4115 MW. The battery is not fully charged because of using the accurate battery charging model that reduces its charging ability at higher state-of-energy. Already in the next time period, the battery operates in the opposite direction. It positions to charge in the DA market, but since it is already almost full and the imbalance settlement price is 40\% higher than the DA price, it is beneficial for the facility to deviate from the DA market position and to discharge the battery, i.e. to sell energy at a price 40\% higher than the DA price. 

Although the battery capacity is 5 MW, it has 9115 MW of the balancing capacity, 4115 MW in the charging and 5 MW in the discharging direction. The battery is mainly used to position in one direction in the DA market and then in reality balance in the other. With the increase of $\Gamma$, the battery tactical positioning is reduced. For $\Gamma=24$ the battery is used strictly for the DA market arbitrage (it charges in hours 4-6, 19, and 22, and discharges in hours 18, 20 and 23) and to balance the facility for different PV scenarios. Fig. \ref{fig:soe} shows that for $\Gamma=0$ the battery constantly operates throughout the day at full power. The exception are the afternoon hours, where it is used to balance the PV output. As $\Gamma$ increases, the zig-zag behavior of the battery state-of-energy is reduced. It is charged during the lowest-price hour 4 and performs arbitrage in the evening between hours 19-20 and 22-23, both pairs having high price differences. Fig. \ref{fig:soe} also shows that the battery requires at least three time periods to fully charge, see e.g. the state-of-energy curves for $\Gamma=12$ and $\Gamma=24$ during hours 4-6.

\subsubsection{Electrolyzer Operation Analysis}

\begin{figure}[t!]
\centerline{\includegraphics[width= .5\textwidth]{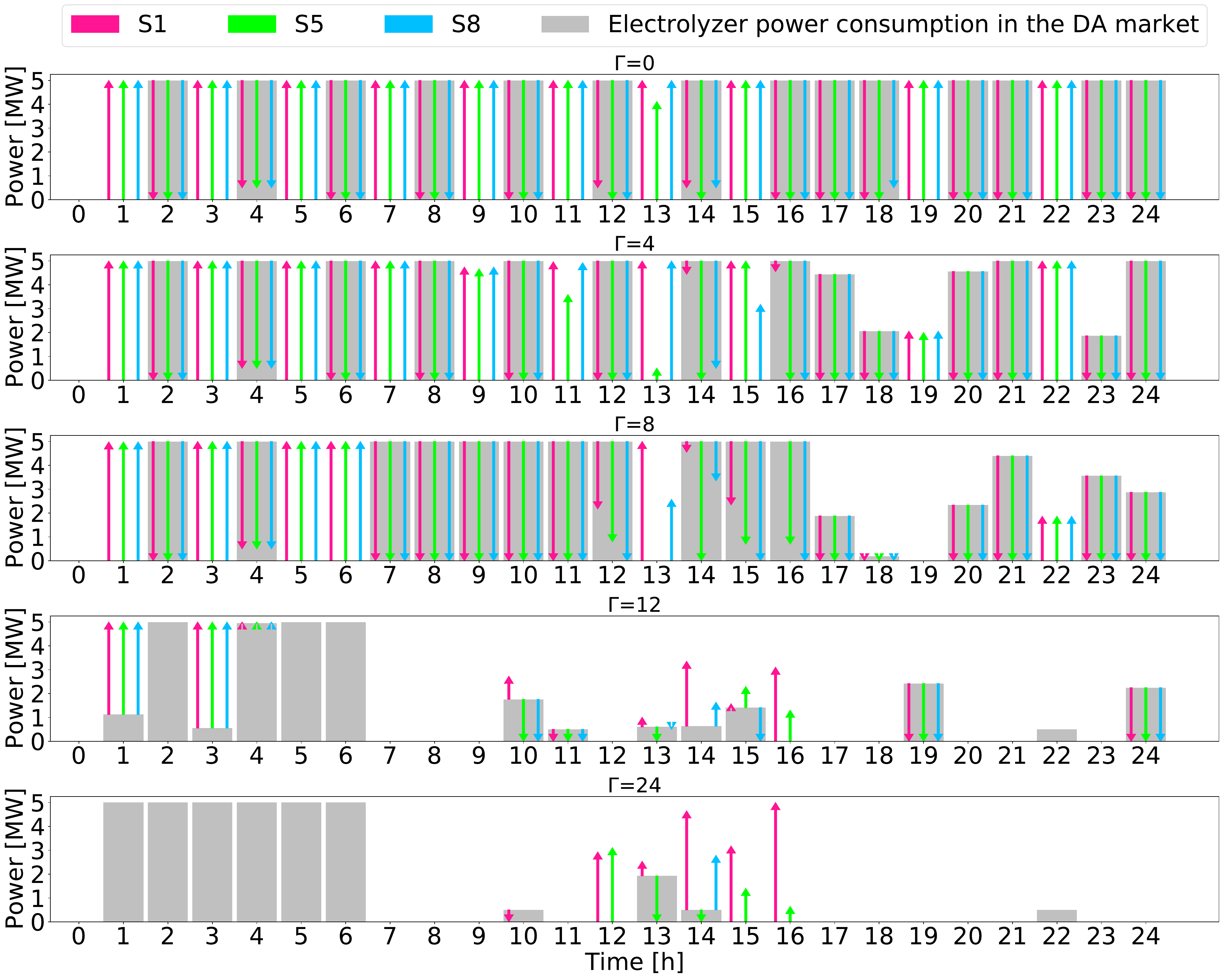}}
\vspace{-3mm}
\caption{Electrolyzer DA schedule and balancing actions for different $\Gamma$ values}
\label{fig:elektrolizator}
\vspace{-4mm}
\end{figure}

\begin{figure}[t!]
\centerline{\includegraphics[width= .5\textwidth]{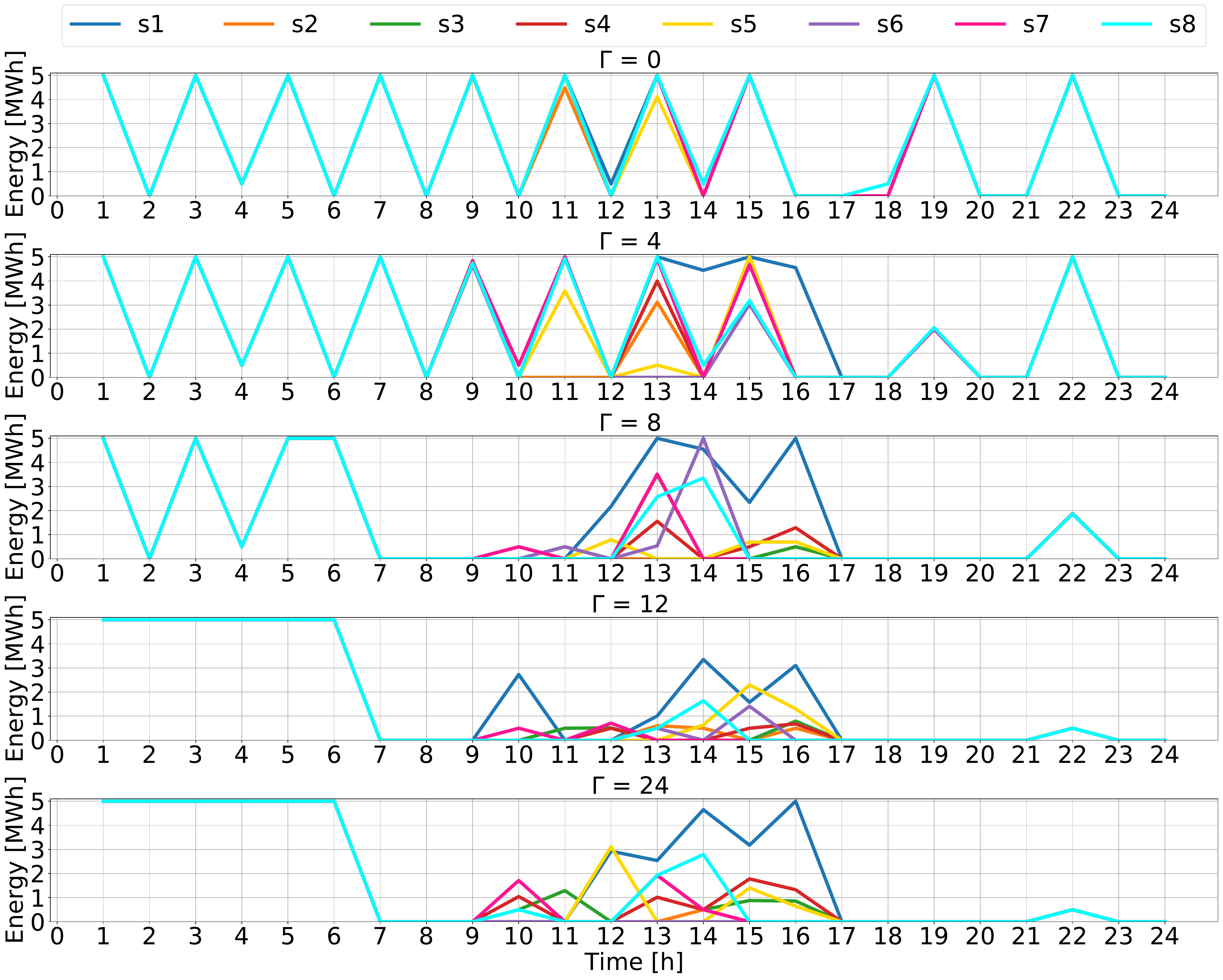}}
\vspace{-3mm}
\caption{Electrolyzer consumption for different $\Gamma$ values for each scenario}
\label{fig:el_cons}
\vspace{-4mm}
\end{figure}

Unlike the battery, the electrolyzer exclusively consumes electricity. In the balancing context, the benefits of this technology are twofold. First, the electrolyzer can contribute to reduction of the imbalance costs and perform arbitrage between the DA market and the imbalance settlement and, second, through its participation in the hydrogen market it can provide revenue to the facility. Fig. \ref{fig:elektrolizator} shows that for $\Gamma=0$ during the first hour the electrolyzer is idle in the DA market. Similar as the battery, due to the expected favorable imbalance price, it is beneficial that the electrolyzer deviates from the DA schedule and operates at its capacity at price 40\% lower than in the DA market. The final outcome is a reduced cost of hydrogen production. In the second hour, the electrolyzer does the opposite, i.e. it positions to consume electricity at the DA stage, but it intentionally deviates and does not consume electricity. This operation is synchronized with the operation of the battery, which dictates the deviation of the entire facility. Thus, a gain from producing hydrogen in hour two is lower than the overall gain of arbitraging between the DA and the balancing stage.
However, in the lowest-price hour 4 the electrolyzer only reduces its output to the technical minimum as the gain from selling hydrogen outperforms the gain from performing arbitrage. Again, as the value of $\Gamma$ increases, the electrolyzer arbitrage between the stages reduces. Ultimately, for $\Gamma=24$ the electrolyzer is used in a conservative manner, it produces hydrogen during the low-price hours and balances the PV output during the day. This is best shown in Fig.  \ref{fig:el_cons}, which presents the electrolyzer operation per scenario.

\vspace{-3mm}
\section{Conclusion} \label{Conclusion}
The paper presents a DA offering model of an RES facility coupled with a battery and an electrolyzer considering the European single-imbalance-pricing scheme. Expectation of the imbalance price, which all European TSOs are required to apply, is modeled through a robust subproblem. The uncertainty of RES production is modeled using stochastic scenarios. The following conclusions are reached: 
\begin{itemize}
    \item In the presented case study both the expected and the actual imbalance revenues are either positive or only slightly negative. This indicates that batteries and electrolyzers have a great potential for imbalance management of RES.
    \item Coupling the power and hydrogen systems is still at the initial stage and the future hydrogen market prices will highly depend on the electricity prices, assuming majority of hydrogen will be produced from green electricity. Use of hydrogen as an energy vector that exploits the surplus of energy and covers the imbalances increases competitiveness of green hydrogen. 
    The presented model can easily be applied to future EU systems.
    \item Applying the robust approach to a bidding structure with consecutive markets results in aggressive arbitrage between those markets. However, the final economical outcome for the facility is strongly dependent on the power system deviation directions and the results obtained for low values of $\Gamma$ should be taken with consideration.
    \item The authors are aware that such behavior is generally illegal and the purpose of this paper is to provide a warning that a possibility of speculative bidding exists and deserves to be addressed. While the main idea behind the single-imbalance-pricing scheme is to reward those who help the system in real time, the presented study indicates there is a possibility that flexible assets intentionally cause imbalance. Seeking opportunities through arbitrage between the DA energy market and the imbalance settlement is generally not allowed and one of the purposes of this paper is to provide a warning that hybrid facilities can be easily engaged in such speculative trading. While the main idea behind the single-imbalance-pricing scheme is to reward those who help the system in real time, the presented study indicates there is a possibility that flexible assets intentionally cause imbalances if they can lead to lower costs/higher revenues. This issues should be further assessed from the policy and market monitoring standpoint and relevant authorities, such as European national energy regulators and the EU Agency for the Cooperation of Energy Regulators, should closely track the behavior of new market participants. 
\end{itemize}    
The future research direction will focus on the investment problem, with the aim of achieving the desired return periods considering the incentives for producing renewable hydrogen.

\vspace{-6mm}
\bibliography{Uravnotezenje_cse}
\bibliographystyle{IEEEtran}

\end{document}

%% file: _shorthands.tex
\newcommand{\omegaI}{\Omega^\mathrm{I}_{}}

\newcommand{\balprice}{\lambda^\mathrm{B}_{t}}
\newcommand{\DAprice}{\lambda^\mathrm{DA}_{t}}
\newcommand{\IDprice}{\lambda^\mathrm{ID}_{t}}
\newcommand{\windplanned}{RES^\mathrm{p}_{t}}
\newcommand{\windreal}{RES^\mathrm{r}_{t,s}}
\newcommand{\batprice}{\lambda^\mathrm{bc}_{}}
\newcommand{\powprice}{\lambda^\mathrm{bp}_{}}
\newcommand{\eff}{\eta^\mathrm{}_{}}
\newcommand{\R}{R^\mathrm{}_{i}}
\newcommand{\F}{F^\mathrm{}_{i}}
\newcommand{\soemaxpar}{\overline{SOE_{}}}
\newcommand{\pmaxpar}{\overline{P^\mathrm{grid}_{}}}
\newcommand{\pmax}{\overline{P^\mathrm{bat}_{}}}

\newcommand{\Hprice}{\lambda^\mathrm{H}_{t}}
\newcommand{\Wprice}{\lambda^\mathrm{W}_{t}}
\newcommand{\elmax}{\overline{P^\mathrm{el}_{}}}
\newcommand{\eleff}{\eta^{\mathrm{w}}}
\newcommand{\alfa}{\alpha}
\newcommand{\bet}{\beta}
\newcommand{\elCoef}{\vartheta^{\mathrm{h}}}
\newcommand{\elmin}{\zeta^{\mathrm{h}}}
\newcommand{\coef}{\kappa^{\mathrm{B}}}
\newcommand{\eBD}{ib^{\mathrm{d}}_{t}}
\newcommand{\eBS}{ib^{\mathrm{s}}_{t}}

\newcommand{\devplus}{res^\mathrm{+}_{t}}
\newcommand{\devminus}{res^\mathrm{-}_{t}}
\newcommand{\devplusmax}{\overline{res^\mathrm{+}_{t}}}
\newcommand{\devminusmax}{\overline{res^\mathrm{-}_{t}}}

\newcommand{\realv}{res^\mathrm{r}_{t,s}}
\newcommand{\dev}{d_{t,s}}
\newcommand{\chDA}{ch^\mathrm{DA}_{t}}
\newcommand{\chID}{ch^\mathrm{ID}_{t}}
\newcommand{\disDA}{dis^\mathrm{DA}_{t}}
\newcommand{\disID}{dis^\mathrm{ID}_{t}}
\newcommand{\soemax}{soe^\mathrm{max}_{}}
\newcommand{\real}{r_{t,s}}
\newcommand{\markpos}{mp_{t}}
\newcommand{\soe}{soe^\mathrm{}_{t}}
\newcommand{\soes}{soe^\mathrm{}_{t,s}}
\newcommand{\deltasoe}{\Delta soe^\mathrm{}_{t}}
\newcommand{\deltasoes}{\Delta soe^\mathrm{}_{t,s}}
\newcommand{\deltat}{\Delta t^\mathrm{}_{}}
\newcommand{\soet}{soe^\mathrm{}_{t-1}}
\newcommand{\soets}{soe^\mathrm{}_{t-1,s}}
\newcommand{\xch}{x^\mathrm{ch}_{t}}
\newcommand{\xdis}{x^\mathrm{dis}_{t}}
\newcommand{\xchs}{x^\mathrm{ch,EU}_{t}}
\newcommand{\xdiss}{x^\mathrm{dis,EU}_{t}}

\newcommand{\soepom}{ soe^\mathrm{p}_{t,s,j}}
\newcommand{\soepomt}{ soe^\mathrm{p}_{t-1,s,j}}
\newcommand{\Rplusi}{ R^\mathrm{}_{i+1}}
\newcommand{\Rminusi}{ R^\mathrm{}_{i-1}}
\newcommand{\Fjedan}{ F^\mathrm{}_{i=1}}
\newcommand{\Fplusi}{ F^\mathrm{}_{i+1}}
\newcommand{\Fminusi}{ F^\mathrm{}_{i-1}}

\newcommand{\elDA}{el^\mathrm{DA}_{t}}
\newcommand{\hyd}{\chi^{\mathrm{h}}_{t,s}}
\newcommand{\ph}{el_{t,s}} 
\newcommand{\eh}{e^{\mathrm{h}}_{t}} 
\newcommand{\phcrtano}{\widehat{el}_{t,s}} 
\newcommand{\binH}{x^{\mathrm{e}}_{t}}
\newcommand{\binHdvaplus}{x^{\mathrm{e,B+}}_{t,s}}
\newcommand{\binHdvaminus}{x^{\mathrm{e,B-}}_{t,s}}
\newcommand{\binB}{x^{\mathrm{b,B}}_{t,s}}
\newcommand{\binDA}{x^{\mathrm{b,DA}}_{t}}
\newcommand{\binDual}{b_{t,s}}

\newcommand{\chEUplus}{ch^\mathrm{B+}_{t,s}}
\newcommand{\chEUminus}{ch^\mathrm{B-}_{t,s}}
\newcommand{\disEUplus}{dis^\mathrm{B+}_{t,s}}
\newcommand{\disEUminus}{dis^\mathrm{B-}_{t,s}}
\newcommand{\elEUplus}{el^\mathrm{B+}_{t,s}}
\newcommand{\elEUminus}{el^\mathrm{B-}_{t,s}}

\newcommand{\dualvar}{\omega_{s}}
\newcommand{\dualvardva}{z_{t,s}}
\newcommand{\dualvartri}{y_{t,s}}
\newcommand{\dualfeas}{\mu_{1,t}}
\newcommand{\dualfeasdva}{\mu_{2,t}}
\newcommand{\bigM}{M}
\newcommand{\binDualFeas}{x^{\mathrm{d}}_{1,t}}
\newcommand{\binDualFeasdva}{x^{\mathrm{d}}_{2,t}}

%% file: diagram.tikz
\begin{tikzpicture}
\node[draw,rounded corners = 2pt, fill=blue!20,very thick, rectangle, minimum width=5cm, minimum height=1cm, anchor=south west, text width=7cm, align=center] (A) at (0,0) {\footnotesize \textbf{Optimization problem:}  \\ Maximization of the hybrid facility's profit};

\node[draw,rounded corners = 2pt, fill=blue!20,very thick, rectangle, minimum width=5cm, minimum height=1cm, anchor=south west, text width=7cm, align=center] (B) at (0,-2) {\footnotesize \textbf{Robust subproblem:} \\ Selection of the worst deviation
directions of the system, respecting the uncertainty budget };

\draw[-{Stealth[length=3mm]},very thick] ($(A.270) + (-1,0)$) -- node[left] {\footnotesize $ b_{t,s} $} ($(B.90) + (-1,0)$);

\draw[-{Stealth[length=3mm]},very thick] ($(B.90) + (1,0)$) -- node[right] {\footnotesize $ \omega_{s} $, $ z_{t,s} $, $ y_{t,s} $} ($(A.270) + (1,0)$); 

\end{tikzpicture}

%% file: diagram2.tikz
\begin{tikzpicture}
\node[draw,rounded corners = 2pt, fill=blue!20,very thick, rectangle, minimum width=0.5cm, minimum height=1.cm, anchor=south west, text width=1.3cm, align=center] (A) at (0,0) {\footnotesize \textbf{Original\\ OF (1)} };

\node[draw,rounded corners = 2pt, fill=blue!20,very thick, rectangle, minimum width=1cm, minimum height=1cm, anchor=south west, text width=2cm, align=center] (B) at (2.2,0) {\footnotesize \textbf{Initial robust\\ formulation (2)}};

\node[draw,rounded corners = 2pt, fill=blue!20,very thick, rectangle, minimum width=1cm, minimum height=1cm, anchor=south west, text width=2cm, align=center] (C) at (5.2,-0.15) {\footnotesize \textbf{Dual robust\\ formulation (3)--(6)}};

\node[draw,rounded corners = 2pt, fill=blue!20,very thick, rectangle, minimum width=1cm, minimum height=1cm, anchor=south west, text width=2cm, align=center] (D) at (8,-0.15) {\footnotesize \textbf{Final robust\\ formulation (7)--(17)}};

\draw[-{Stealth[length=3mm]},very thick] ($(A.0) + (0,0)$) -- node[below,yshift=-0.8cm, align=center,text width=2.3cm] {\footnotesize Robustify  $\DAprice \!\cdot\! (1 \!\pm\! \coef)\! \cdot\!  \dev$} ($(B.180) + (0,0)$);

\draw[-{Stealth[length=3mm]},very thick] ($(B.0) + (0,0)$) -- node[below,yshift=-0.8cm, align=center,text width=2.5cm] {\footnotesize Transform robust subproblem into dual}($(C.180) + (0,0)$);

\draw[-{Stealth[length=3mm]},very thick] ($(C.0) + (0,0)$) -- node[below,yshift=-0.8cm, align=center,text width=4cm] {\footnotesize Reduce optimization to single level and reformulate absolute value} ($(D.180) + (0,0)$);

\end{tikzpicture}